\documentclass[letterpaper, 10 pt, conference]{IEEEtran}  

\IEEEoverridecommandlockouts                              
\usepackage{soul}
\usepackage{hyperref}
\usepackage[utf8]{inputenc}
\usepackage[T1]{fontenc}
\usepackage{graphicx}
\graphicspath{ {./images/} }
\usepackage{float}
\usepackage{lipsum}  
\usepackage{array}
\usepackage{multirow}
\usepackage{hyperref}
\usepackage{xcolor}
\usepackage[tikz]{bclogo}
\usepackage[framemethod=tikz]{mdframed}
\usepackage[many]{tcolorbox}
\usepackage{booktabs,xcolor,siunitx}

\usepackage[sorting=none]{biblatex}
\usepackage{collcell}
\usepackage{hhline}
\usepackage{pgf}
\usepackage{amsmath,amssymb,amsfonts}

\newcommand\gray{gray}
\newcommand{\bmm}[1]{\ensuremath{\mathbf{#1}}}

\usepackage{subfigure}
\usepackage{caption}

\newcommand\ColCell[1]{%
  \pgfmathparse{#1<.8?1:0}%
    \ifnum\pgfmathresult=0\relax\color{white}\fi
  \pgfmathparse{1-#1}%
  \expandafter\cellcolor\expandafter[%
    \expandafter\gray\expandafter]\expandafter{\pgfmathresult}#1}

\newcolumntype{E}{>{\collectcell\ColCell}c<{\endcollectcell}}

\newcolumntype{P}[1]{>{\centering\arraybackslash}p{#1}}

\newcolumntype{M}[1]{>{\centering\arraybackslash}m{#1}}

\title{\LARGE \bf
Many-to-One Knowledge Distillation of Real-Time Epileptic Seizure Detection for Low-Power Wearable Internet of Things Systems
}

\author{Saleh Baghersalimi$^{1}$, Alireza Amirshahi$^{1}$, Farnaz Forooghifar$^{1}$, Tomas Teijeiro$^{1}$, Amir Aminifar$^{2}$, David Atienza$^{1}$
\thanks{$^{1}$Ecole Polytechnique Fédéral de Lausanne (EPFL), Switzerland, Email: \texttt{saleh.baghersalimi@epfl.ch}, \texttt{alireza.amirshahi@epfl.ch}, \texttt{farnaz.forooghifar@epfl.ch}, \texttt{tomas.teijeiro@epfl.ch}, \texttt{david.atienza@epfl.ch}}
\thanks{$^{2}$Lund University, Sweden, Email: \texttt{amir.aminifar@eit.lth.se}}
}

\newcommand\MyBox[2]{
  \fbox{\lower0.75cm
    \vbox to 1.7cm{\vfil
      \hbox to 1.7cm{\hfil\parbox{1.4cm}{#1\\#2}\hfil}
      \vfil}%
  }%
}

\begin{document}
\maketitle
\thispagestyle{empty}
\pagestyle{empty}

\begin{abstract}

Integrating low-power wearable Internet of Things (IoT) systems into routine health monitoring is an ongoing challenge. Recent advances in the computation capabilities of wearables make it possible to target complex scenarios by exploiting multiple biosignals and using high-performance algorithms, such as Deep Neural Networks (DNNs). There is, however, a trade-off between performance of the algorithms and the low-power requirements of IoT platforms with limited resources. Besides, physically larger and multi-biosignal-based wearables bring significant discomfort to the patients. Consequently, reducing power consumption and discomfort is necessary for patients to use IoT devices continuously during everyday life. To overcome these challenges, in the context of epileptic seizure detection, we propose a many-to-one signals knowledge distillation approach targeting single-biosignal processing in IoT wearable systems. The starting point is to get a highly-accurate multi-biosignal DNN, then apply our approach to develop a single-biosignal DNN solution for IoT systems that achieves an accuracy comparable to the original multi-biosignal DNN. To assess the practicality of our approach to real-life scenarios, we perform a comprehensive simulation experiment analysis on several state-of-the-art edge computing platforms, such as Kendryte K210 and Raspberry Pi Zero.

\end{abstract}


\section{Introduction}

Epilepsy is amongst the five most common chronic diseases and, according to WHO, it is the most common chronic brain disease affecting more than 50 million people of all ages~\cite{Epilepsy_online}. Besides suffering an associated stigma and discrimination, epilepsy represents the second neurological cause of years of potential life loss. The most life-threatening effect of seizure attacks is sudden unexpected death in epilepsy (SUDEP)~\cite{Lancet}. To reduce the aforementioned effects of such attacks, we need to continuously monitor these patients for life-long periods. With the widespread adoption of Internet of Things (IoT)~\cite{al2015internet} in electronic medical care, low-power and easy-to-use IoT devices are growing in popularity. Thus, we can perform long-term everyday patient monitoring~\cite{ryvlin2020seizure} to inform their family members or caregivers in case of emergency situations.

\begin{figure}[t]
	\begin{center}
	\centering
		\includegraphics[scale=0.51]{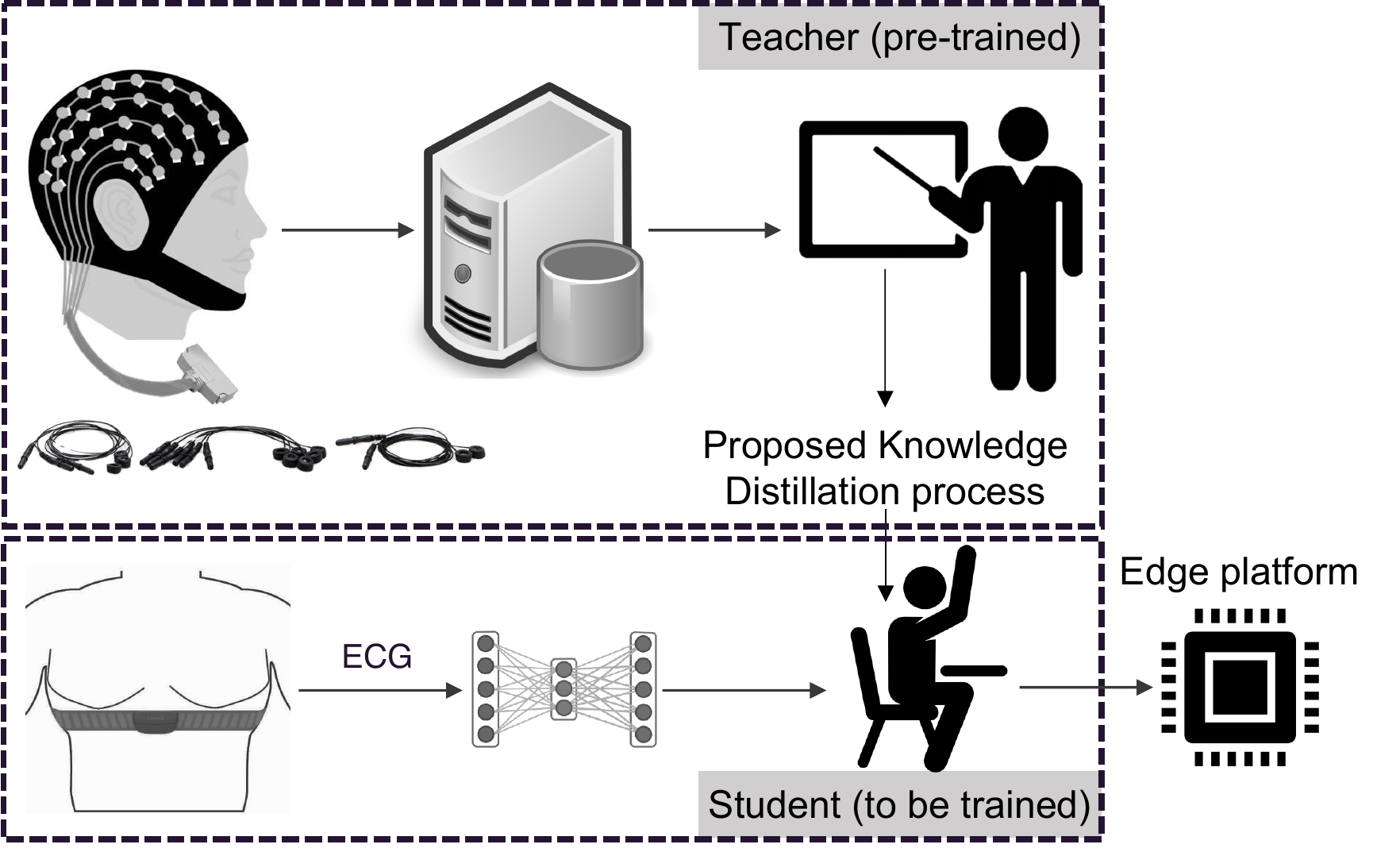}
		\caption{Structure of our proposed knowledge distillation framework for IoT wearable systems. Our student model requires only the ECG signal and can be applied on low-power wearable IoT platforms to perform epileptic seizure detection. In the student model, the patient employs just one sensor, which is comfortable, causes no social stigma, and achieves the detection accuracy of the more complex teacher model.}
		\label{intro_overall}
	\end{center}
\end{figure}

 The current trend in pathology detection systems are based on deep neural networks (DNNs)~\cite{miotto2018deep}, which are capable of classifying large volumes of biomedical data. The ability of these networks to extract high-level and complex patterns from biomedical signals makes them an attractive tool for epilepsy monitoring~\cite{gabeff2021interpreting}. The IoT devices typically own limited resources and cannot satisfy the computing requirement of complex DNNs. By addressing the challenges regarding the comfortability, complexity and energy consumption, these systems can perfectly fit the requirements of epileptic seizure detection systems~\cite{acharya2018deep}. This can be translated in using simpler and lighter networks, preferably with a smaller number of acquired biosignals, to make them more comfortable and improve their energy efficiency regarding data acquisition and processing. 

According to literature, monitoring the electroencephalogram (EEG) signal of the patients provides the standard detection accuracy in case of epileptic seizures~\cite{vidyaratne2017real}. However, by simultaneous monitoring other bio-signals, the decision can become more accurate and robust, as each signal can reflect different effects of the seizures~\cite{leijten2018multimodal}. In this paper, we first introduce our multi-modal detection IoT system, which uses separate residual neural networks~\cite{7780459} for different biosignals, namely, EEG and electrocardiogram (ECG) signals.   

The overhead that comes with this multi-modal bio-signal monitoring is an increase in size of the detection system, as we should use a separate network for each signal. This bigger size results in higher energy consumption~\cite{kim2015compression}, which is problematic for IoT wearable systems and for long-term monitoring purposes. Besides, the synchronisation of signals from multiple devices is challenging in real IoT deployments, thus working with one signal has less risks of incorrect annotation. Finally, connecting several wearable devices to the patient increases the stigma especially in the case of EEG, where devices are usually bulky and draw a lot of attention~\cite{van2016non}.

To reduce the abovementioned overheads caused by multi-modal monitoring, we propose to exploit the concept of knowledge distillation~\cite{gou2020knowledge} in IoT systems. This also helps to leverage the information of both biosignals to keep a satisfying detection quality. Knowledge distillation is a procedure for model compression, in which a small (student) model is trained to match a large pre-trained (teacher) model. In the simplest form of distillation, knowledge is transferred to the distilled model by training it on a transfer set, trying to maximize the cross entropy of the two models. At the same time, the system attempts to maximize the generation of correct labels for the transfer set in the student model~\cite{hinton2015distilling}. Following this concept, as shown in Fig.~\ref{intro_overall}, in this work, we propose a new approach based on knowledge distillation to design high-precision and low-power wearable systems for epileptic seizure detection. Thus, we develop and use a pre-trained multi-model DNN using synchronized EEG and ECG as the teacher and train a single input network using the ECG exclusively as biosignal detection input, while achieving a comparable performance and targeting low-power embedded devices. In the teacher model, which is a complex model with high energy consumption, the patient uses multiple IoT sensors, which are not comfortable and cause stigma. Conversely, the student model contains a much simpler model resulting in less energy consumption and enables more prolonged monitoring with IoT devices, as it requires only the ECG signal. This signal can be acquired using a chest strap, which is comfortable to use and fully removes the social stigma. At the same time, the student model achieves the detection accuracy of the more complex teacher model. To the best of our knowledge, this work is different from the usual knowledge distillation frameworks that have been proposed in the IoT systems context in the sense that other works propose to change the system structure, but the inputs are the same. On the contrary, we are completely omitting the EEG in the student model and only require ECG inputs in our final IoT system design. 

Thus, the main contributions of our work are as follows:

\begin{itemize}
    \item {We develop a knowledge distillation-based approach to develop high-precision low-power wearable IoT systems for epileptic seizure detection in real-time using only a single biosignal input. We show a way to create a neural network trained from the knowledge of a multi-modal DNN system relying on both EEG and ECG signals; but only using the ECG signal as input in the detection phase during the use of the IoT system to detect epileptic seizures. As a result, the synchronized EEG signal is only used in the initial (off-line) training phase. This approach results in removing the complexity and power requirements of EEG acquisition to better fit the design constraints of low-power IoT wearable devices. Moreover, we reduce the stigma and inconvenience of synchronising multiple biosignals in real-life IoT system deployments.}
 
    \item{As initial teacher network for our approach, we design a new multi-modal and multi-channel seizure detection DNN that increases the quality of seizure detection by acquiring information from both EEG and ECG. In this first step, to limit the design complexity and highlight how to target different biosignal inputs, we conceive independent 1-dimensional DNNs for each signal. Then, we use a weighted combination of the results to reach a final decision. Compared with a network using exclusively the ECG signal for training, we show that the proposed multi-modal method increases the sensitivity by 6.12\% and specificity by 16.07\% when applied to the EPILEPSIAE dataset~\cite{Ihle:2012:EEE:2207274.2207331}. Our approach increases seizure detection accuracy and reduces false alarms at the same time, in comparison with using only the ECG signal. }

    \item{By using knowledge distillation, we reduce the complexity of the initial teacher DNN system to result in an implementable network for a much lower power IoT wearable device. Our results show the distilled model can be implemented on current IoT wearable platforms. Using an edge AI platform, we show that the required energy capacity is reduced by 37.65\%. This energy reduction is obtained while sensitivity and specificity are only reduced by 1.5\% and 1.3\% in comparison with the initial (and very uncomfortable for the user) multi-modal DNN teacher IoT system.}
\end{itemize}

The rest of this article is organized as follows. In Section~\ref{sec:RelatedWork}, we review previous works on low-power embedded wearable platforms, multi-modal seizure detection and knowledge distillation. In Section~\ref{sec:Methodology}, we present a general overview of our seizure detection systems and the different parts: preprocessing, initial multi-modal network architecture and our knowledge distillation approach targeting low-power IoT wearable systems. Then, Section~\ref{sec:ExperimentalSetup} presents our experimental setup and, in Section~\ref{sec:Result}, we analyze the computational and energy consumption characteristics of the proposed distilled implementation, and compare it with other similar IoT wearable architectures. Finally, in Section~\ref{sec:Conclusion}, we summarize the main conclusions of this work.
\section{Related Work}
\label{sec:RelatedWork}

The gold standard for non-invasive seizure detection is EEG monitoring~\cite{shellhaas2015continuous}, which has been used for decades in highly specialized and costly hospital environments. EEG-based seizure detection has received noticeable attention in the literature as brain activity is significantly affected during seizure attacks. In~\cite{fan2018detecting}, the authors proposed the use of spectral graphs to extract spatial-temporal patterns for seizure detection. In~\cite{wang2018epileptic}, they used the wavelet transform, which has been applied to the time-frequency domain for the detection of epileptic activity. In~\cite{samiee2014epileptic}, the authors proposed an effective feature extraction algorithm named discrete short-time Fourier transform (DSTFT), which is an adaptive generalization of the classical short-time Fourier transform (STFT). In \mbox{\cite{olokodana2021ezcap}}, the authors apply the Kriging methods on EEG signals in a wearable system configuration to reduce the latency in real-time epileptic seizure detection. In \mbox{\cite{olokodana2020real}}, the same Kriging methods have been used for the application of early seizure detection. The authors in \mbox{\cite{sayeed2019neuro}} use discrete wavelet transform and statistical features to apply preprocessed EEG signals to a neural network classifier to detect epileptic seizures. 
All the works mentioned above propose power-hungry algorithms that use a complete set of EEG channels. Therefore, if these methods are applied to real-life IoT  wearable systems, patients would suffer from the social stigma of wearing a cap with electrodes on the head. Yet, our work presents a knowledge distillation to remove the most invasive signals used in continuous monitoring, leading to a more straightforward, more energy-efficient, and less stigmatizing setup applicable to IoT wearable systems. 


To optimize energy consumption, in~\cite{pullini2019mr} a hierarchical architecture with a wide set of near-sensor processing kernels is presented. They have shown that, due to their power management techniques, their architecture is suitable for IoT wearable systems. 
In~\cite{de2020modular}, the authors have used optimization and parallelization techniques besides integration of domain-specific accelerators with the goal of improving the mapping and reducing the energy consumption of biomedical applications. The technique that we are targeting here is knowledge distillation, which is originally a model compression technique. This enables us to train smaller and less complex DNNs with comparable performance. To interpret DNNs for epileptic seizure detection on EEG signals, in~\cite{gabeff2021interpreting}, they associate certain properties of the model behavior with the expert medical knowledge. They have come up with online seizure event characterization able to handle inter-patient variability.

To overcome the social stigma problem in IoT wearable systems using EEG caps, in~\cite{sopic2018glass}, a wearable system based on the temporal EEG electrodes for the detection of epileptic seizures is presented. By reducing the acquired signals to four, the system could be implemented on glasses, which can be easily worn by patients. 
Although these glasses solve the social stigma issue, they use only EEG signals for the seizure detection task, which are hard to obtain compared to other biomedical signals. In contrast, our work uses the knowledge of EEG and ECG signals, while the final model uses only the ECG signal to detect seizures more effortlessly than with EEG signals.

Besides EEG, other biosignals can also get affected by epileptic seizures and provide additional information about the occurrence of such attacks. For example, seizures are often associated with cardiovascular alterations, and measures related to the heart rate are known to be useful clinical signs of an epileptic discharge~\cite{varon2013detection, zijlmans2002heart, leutmezer2003electrocardiographic, forooghifar2019self,forooghifar2019resource}. Thus, combining other biosignals with EEG can result in better detection of different types of seizures, especially those that are not easily detected with just the EEG. Multi-modal seizure detection has been done previously in several works. 
In~\cite{onorati2017multicenter}, a combination of electrodermal activity and accelerometer signals is used. 
In~\cite{furbass2017automatic}, the authors have used EEG, Electromyography (EMG) and ECG signals and have shown that the average sensitivity is improved in comparison with using each individual sensor separately. 
In~\cite{greene2007combination}, a combination of linear discriminant models extracted from EEG and ECG features is used to detect seizures in newborn infants. 
In~\cite{mporas2015seizure}, an SVM model is used with both EEG and ECG signals and could achieve high accuracy being tested on three patients. 
In~\cite{qaraqe2016epileptic}, seizure detection is done using an SVM model on multi-channel EEG and single-channel ECG separately and then fusing them into one final decision. The effect of using this multi-modal method is then shown on the number of false alarms and detection delay. 
Recently, in~\cite{liu2020epileptic}, 
convolutional neural networks (CNN) are used in combination of EEG, ECG and respiration. They have observed that 
the multi-modal system outperforms systems with individual signals. The reduced deep convolutional stack autoencoder is used in another work using 18 channels of EEG signal, which has resulted in very high performance; but yet, the system cannot be implemented on IoT wearable devices and is intrusive~{\cite{sahani2021epileptic}}.


Knowledge distillation in neural networks was first introduced in 2015~\cite{hinton2015distilling}, where a single model is trained from an ensemble of models (known as specialists or experts). The key aspect of knowledge distillation is learning not from the discrete labels of the dataset, but from the continuous output of the "expert" models. This concept has been tested on speech and images primarily but then also showed satisfying results in certain healthcare applications. In that work, it is demonstrated that we can benefit from the knowledge distillation method to efficiently regularize the smaller network and achieve better generalization than directly training the network using the labels of the training set~\cite{chitrakar2016social, howard2017mobilenets}. In~\cite{che2015distilling}, authors have used gradient boosting trees as the experts to train their deep learning model on a real-world clinical time-series dataset, showing an improvement in the performance with respect to their initial deep learning model.

In~\cite{dou2020unpaired}, multi-modal segmentation of computerized tomography (CT) and magnetic resonance imaging (MRI) is compacted by distilling knowledge from cross-modal information of these images. Similarly, in~\cite{li2020dual}, the student model learns from both labeled target data (e.g., CT), and unlabeled target data and labeled source data (e.g., MR) by two teacher models. They have shown that their approach can utilize unlabeled data and cross-modality data with superior performance, outperforming semi-supervised learning and domain adaptation methods with a large margin.  
Based on the reviewed state-of-the-art works, we hypothesise that distilling the knowledge from a multi-modal epileptic seizure detection system to a single-input system, can improve the performance of the single-input network. Thus, in the following section, we will leverage this concept and introduce first our multi-modal system and, after that, our distilled single-input system. 

\section{Proposed Knowledge Distillation Approach}
\label{sec:Methodology}
\begin{figure*}[ht]%
    \begin{center}
	\centering
		\includegraphics[scale=0.41]{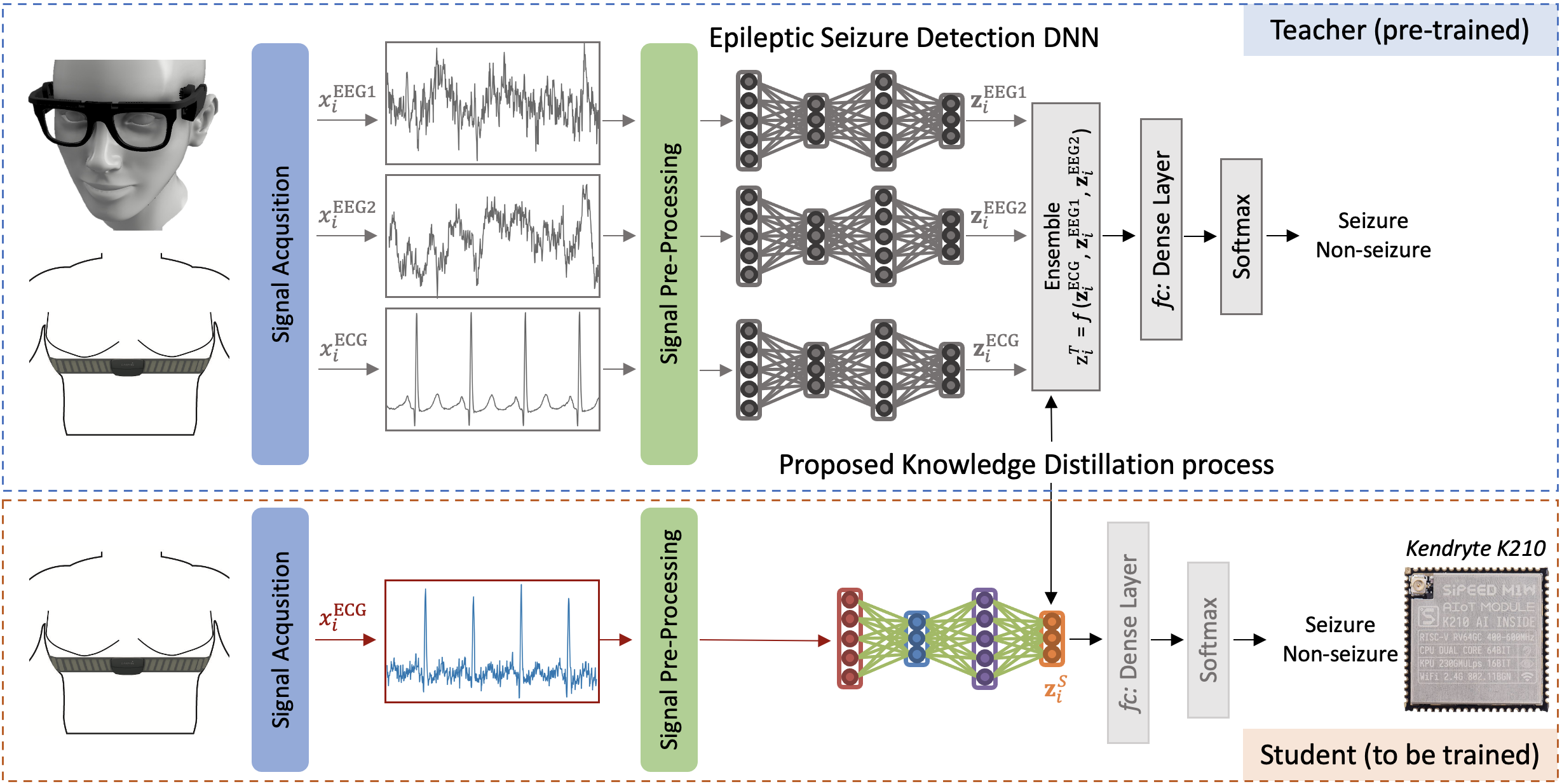}
		\caption{The overall flow of the proposed teacher-student framework for knowledge distillation for epileptic seizure detection. The teacher DNN is an ensemble composed of three Res1DCNN to extract valuable information from the training set. Making epileptic seizure detection using the teacher DNN is heavy to compute as it requires both ECG and EEG signals and is not suitable for IoT wearable systems. The student DNN is a single Res1DCNN that requires only an ECG signal. Thus, it is less computationally intensive. The Kendryte K210, Raspberry Pi Zero system, or the latest multi-core edge AI architectures (e.g., smartwatches or PULP SoC~{\cite{de2020modular}}) perform the epileptic seizure detection using the student DNN on ECG signal acquired by a chest strap.}
		\label{overall}
	\end{center}
\end{figure*}

This section describes the elements of our proposed feature-based knowledge distillation approach for epileptic seizure detection. The overall flow is presented in Fig.~\mbox{\ref{overall}}. Our proposed approach includes a pipeline divided in three phases, namely: signal acquisition and pre-processing, multi-modal CNNs framework, and distilling the knowledge from the more extensive teacher network to the lighter student network. The teacher network requires both ECG and EEG signals and has a more complex network architecture than the student network, which results in more energy consumption. Thus, it is not convenient to be deployed on IoT wearable devices or low-power platforms with limited resources for long-term monitoring. On the other hand, the student network receives only the ECG signal, has fewer parameters, requires lower computational costs, and performs real-time epileptic seizure detection without losing validity. Thus, it is suitable to be deployed in edge platforms and IoT wearable devices.

\subsection{Signal Acquisition and Pre-Processing}
\label{sec:signal_pre_processing}

\begin{figure}[t]
	\begin{center}
	\centering
		\includegraphics[scale=0.67]{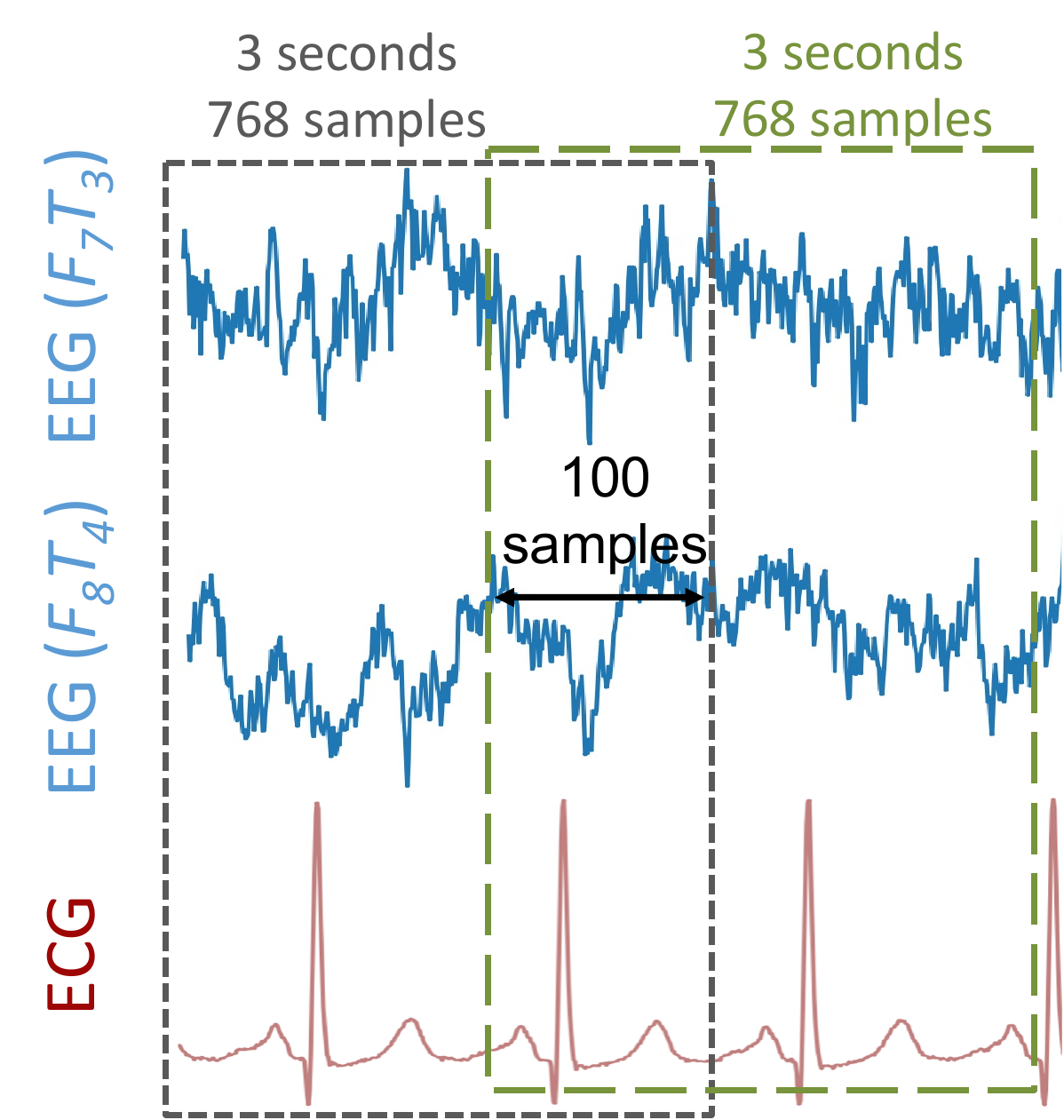}
		\caption{Segmentation of ECG and EEG signals sampled with a frequency of 256 Hz using slots of 3-seconds with 100 samples overlap. Both ECG and EEG signals are synchronized to train the teacher DNN. In other words, both ECG and EEG signals are acquired and measured in parallel. Therefore, we have the corresponding ECG signal when there is a seizure in the EEG signal. Moreover, note that in latest edge AI architectures deployed in the medical IoT ecosystem (e.g., multi-core PULP system~{\cite{de2020modular}}), it is possible to process the signal in less than one second, and during the remaining time the system can be in sleep mode.}  
		\label{signal_segmentation}
	\end{center}
\end{figure}

A DNN requires a considerable amount of data for the training process. The amount of data of actual epileptic seizures for a particular patient is generally limited. Thus, data augmentation techniques have been used to increase the amount and diversity of epileptic data in recent works~\mbox{\cite{pascual2020epilepsygan, zhao2020classification}}. In this work, we exploit data augmentation by segmenting the synchronized ECG and EEG signals acquired by IoT wearable devices with a sampling frequency of 256~Hz. Since the interpretation of the QRS complexes and obtaining their characteristics is one of the essential parts of ECG signal processing, we consider 3-second (768 samples) slots to ensure a minimum of two QRS complexes. These slots are obtained by sliding a fixed-length window with 100 samples overlapping through the entire signal. Figure~\mbox{\ref{signal_segmentation}} shows how the segmentation of ECG and EEG signals is performed in our approach.

\begin{figure*}[t]
	\begin{center}
	\centering
		\includegraphics[scale=0.58]{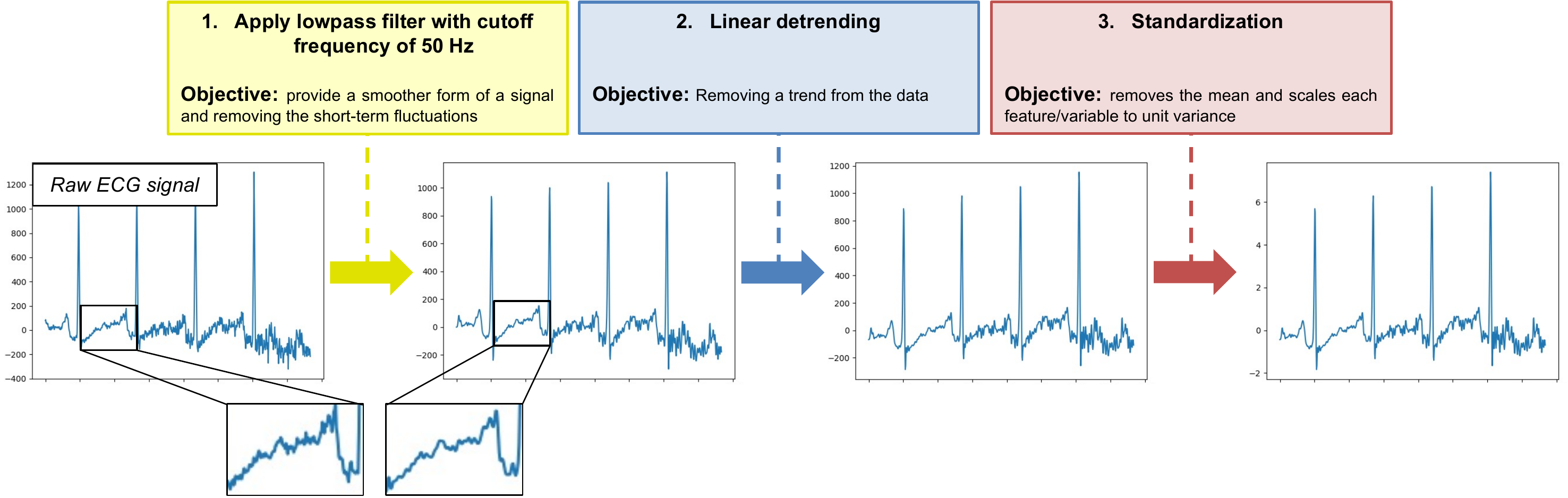}
		\caption{Pre-processing consists of three steps: (1) applying a low-pass filter, (2) Linear detrending to remove linear trend along the axis from data, and (3) standardization to have zero mean and unit variance.}  
		\label{preprocessing}
	\end{center}
\end{figure*}

Pre-processing techniques are required and have been utilized in different applications to train DNN models effectively and efficiently~\cite{huang2020normalization}. Thus, after signal segmentation, we propose a simple method for pre-processing each segment. It consists of three steps as shown in Fig.~\ref{preprocessing}. First, we apply a 10th-order low-pass Butterworth filter with a cutoff frequency of 50~Hz to smooth the signal segment. Secondly, we perform a linear detrending, where the result of a linear least-squares fit to data is subtracted from the initial data. Finally, we apply standardization on each segment to transform it to have zero mean and unit variance~\cite{kessy2018optimal}. The standardization is done on each 3-second segment separately so that a potentially significant difference in the magnitude of one segment does not cause a high degradation in other segments.

\subsection{Epileptic Seizure Detection DNN}
\label{sec:Multi-channel}

This section describes the DNN that is used in this work for seizure detection. We considered our previous work's DNN named as Res1DCNN~\cite{9479691}. 
Let $\{x_{1},...,x_{T}\} : x_{i} \in \mathbb{R}^{r}$, where $r$ denotes the number of samples in a single window. Given a data sample $x_{i}$, Res1DCNN returns $\bmm{z}_i \in \mathbb{R}^{L}$ as a feature map after the convolutional blocks, where $L$ denotes the length of the feature map. A simple fully-connected layer transforms $\bmm{z}_i$ to $\hat{y}_{i}$ which predicts the label of the data sample, i.e., seizure or non-seizure.
%
 This model is chosen for our system since it can produce promising results in seizure detection without being computationally intensive.
%
Moreover, Res1DCNN is efficient in energy consumption and detection performance on mobile platforms with tight energy operation constraints (as discussed in Section~\ref{sec:Result}). 

\begin{figure}[t]
	\begin{center}
	\centering
		\includegraphics[scale=0.31]{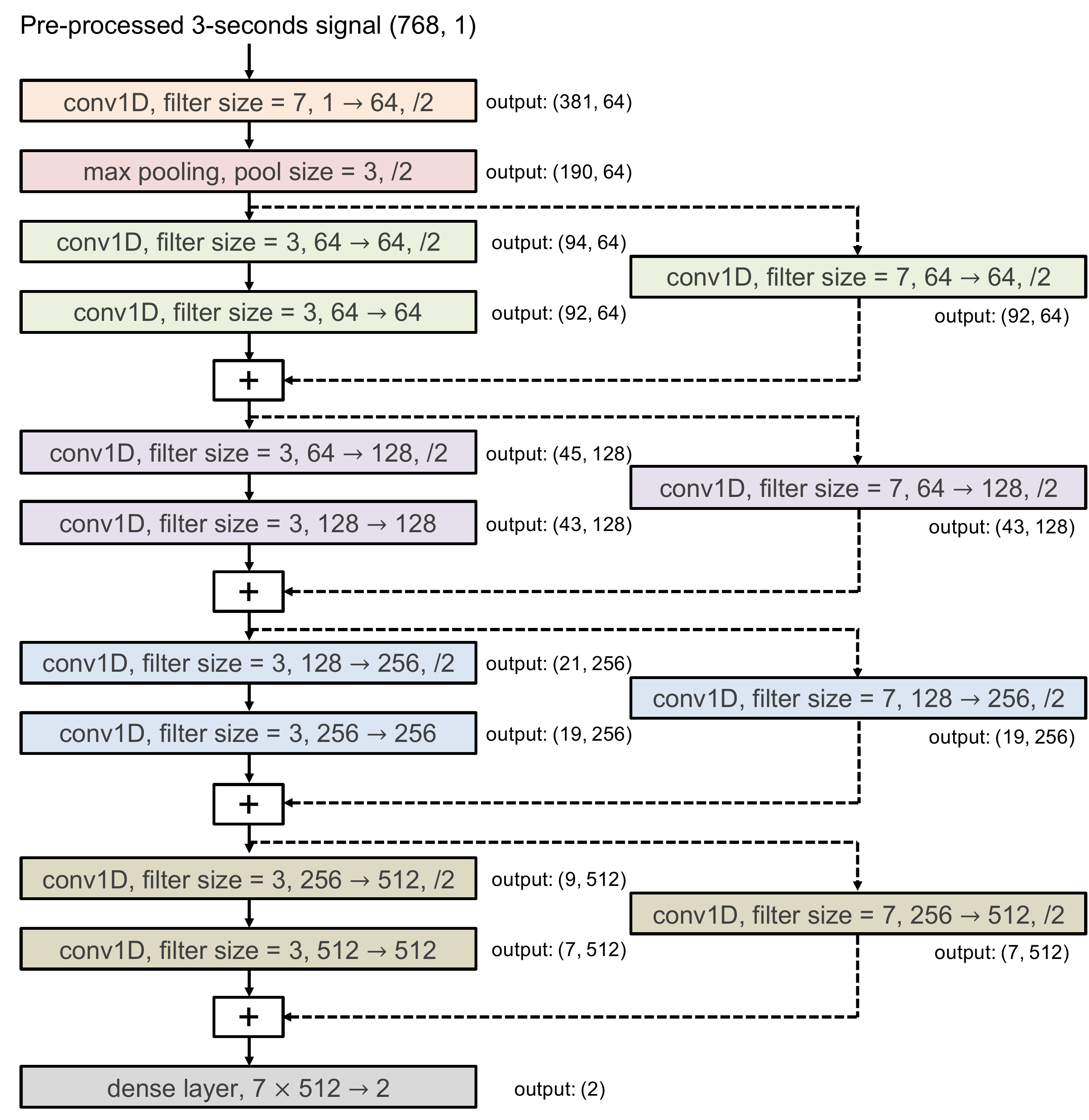}
		\caption{The architecture of the Res1DCNN~\cite{9479691}. The network contains 14 convolutional layers with skip connections followed by a dense layer. Here, `/2' denotes the downsampling operator using a strided convolution with a factor of 2. `$\rightarrow$' denotes the transition from the input to output channels.}
		\label{res1dcnn_figure_architecture}
	\end{center}
\end{figure}

The architectural details of our proposed end-to-end Res1DCNN model are shown in Fig.~\ref{res1dcnn_figure_architecture}. Res1DCNN includes multiple skip connections similar to those found in Residual Neural Networks~\cite{7780459}. These skip connections make residual blocks norm-preserving, allowing us to propagate the information in very deep neural networks to render the training stable. The resulting network consists of only 14 weight layers, making it appropriate to be implemented on a battery-powered edge AI platform with limited computational resources.

\subsection{Teacher Network: Multi-modal Res1DCNN}
\label{sec:teacher}
In this work, the teacher network uses the Res1DCNN architecture. The teacher takes synchronized ECG and 2-channel EEG segments as inputs and trains three analogous Res1DCNN models for every input.
This method is an ensemble learning technique where various DNN models are combined and trained to solve the same problem~\mbox{\cite{zhang2012ensemble}}. 
As shown in Section~\mbox{\ref{detection_performance_analysis}}, by using our proposed ensemble learning method named Multi-modal Res1DCNN, we have a more accurate DNN model to perform epileptic seizure detection. We could use all EEG channels; however, since we studied and experimented with an IoT wearable system setup, we considered only two EEG channels where the signals can be acquired from e-Glass~\mbox{\cite{sopic2018glass}}, as shown in Fig.~\mbox{\ref{eeg_electrodes}}.

In this network, the inputs are EEG and ECG segments $\{x^{\text{ECG}}_i, x^{\text{EEG1}}_i, x^{\text{EEG2}}_i\}$. We extract features from the inputs by passing each of them through Res1DCNN's feature extractor to obtain $\{\bmm{z}^{\text{ECG}}_i, \bmm{z}^{\text{EEG1}}_i, \bmm{z}^{\text{EEG2}}_i \}$. We merge these feature maps into a single $\bmm{z}_i^{T} \in \mathbb{R}^{L}$ using a linear combination of the features. More formally, $\bmm{z}_i^{T} = f(\bmm{z}^{\text{ECG}}_i, \bmm{z}^{\text{EEG1}}_i, \bmm{z}^{\text{EEG2}}_i ; \theta)$, where $\theta$ is the trainable weight for the linear combination. When $\bmm{z}_i^{T}$ is obtained, we train a simple fully-connected layer that predicts the output $\hat{y}_i$ from $\bmm{z}_i^T$. Finally, a softmax layer outputs the predicted value. The softmax layer is a generalization of the logistic layer that highlights the largest values in a vector, while suppressing the values that are significantly below the maximum.


\subsection{Student Network: Distilling the Knowledge}
\label{sec:distilling_the_Knowledge}

In machine learning (ML) methods, training an ensemble of various models using the same data is a solution to improve performance~\cite{dietterich2000ensemble}. However, making predictions using most DNNs and ensemble models requires significant storage and is too computationally expensive. Consequently, as our goal is to implement the epileptic seizure detection algorithm on an IoT wearable medical platform that runs on a battery, the teacher network is not an appropriate model for implementation. Moreover, as described in Section~\ref{sec:teacher}, the teacher DNN uses different input data, such as EEG and ECG. However, in a real-world scenario, acquiring EEG signal is complex and uncomfortable for the user. In this work, to address the problems mentioned above, we introduce a student network that gets only ECG signals as its input. As demonstrated in~\mbox{\cite{buciluǎ2006model}}, it is achievable to compress the knowledge from an ensemble model into a single student model. This process enables the model to run on embedded devices, considering these devices' stringent energy and memory constraints.

In this work, as Fig.~\ref{overall} shows, we employ knowledge distillation, a model compression method where a small DNN is trained to simulate a pre-trained larger DNN. This training is referred to as "Teacher-Student", where the large DNN is the teacher network, and the small DNN is the student network~\cite{hinton2015distilling, kim2018paraphrasing, ba2013deep}. 
In our proposed framework, the student network is a Res1DCNN that gets only $x^{\text{ECG}}_i$ in the input, and the output of the model is $\bmm{z}^S_i \in \mathbb{R}^{L}$. To transfer the knowledge from the pre-trained teacher into the student, we define the loss function as an L2 distance between $\bmm{z}^S_i$ and $\bmm{z}^T_i$. Training the model with this loss function causes the feature map of the student to be similar to the feature map of the teacher. Therefore, this similarity enables us to use the fully-connected layer of the teacher network in the student network without retraining. 
As defined in the loss function, we use the feature map $z^T_i$ in the teacher model because it has a superior signal intensity and spatial correlation information. 
 
  Knowledge distillation, in this work, enables us to use only ECG segments as the input for our final IoT wearable system in real-life operation. Thus the student network only processes the ECG inputs, and the Res1DCNN is instantiated only once, while in the teacher, we have three parallel Res1DCNN models. Furthermore, as a consequence, the amount of computation is considerably decreased while the network's performance is maintained. Furthermore, using the student network, we can translate the proposed application to a real-life scenario that can benefit patients, clinicians, etc.  This is done by removing the necessity to permanently wear a cap to monitor EEG outside the hospital, which causes social stigma and discomfort for patients.


\section{Experimental Setup}
\label{sec:ExperimentalSetup}
In this section, we present the experimental setup to evaluate our proposed feature-based knowledge distillation approach in terms of epileptic seizure detection performance and energy consumption.

\subsection{Epileptic Seizures Dataset}

For our experiments, we use the EPILEPSIAE dataset~\cite{Ihle:2012:EEE:2207274.2207331}, which is one of the most extensive epilepsy datasets manually annotated by medical experts for seizure detection and prediction worldwide. This dataset is made in a routine clinical environment and consists of one-lead ECG and 19-channel EEG data of 30 patients. No constraints regarding the types of seizures are imposed; the dataset contains complex partial, simple partial, and secondarily generalized seizure types.

\begin{figure}[t]
	\begin{center}
	\centering
		\includegraphics[scale=0.26]{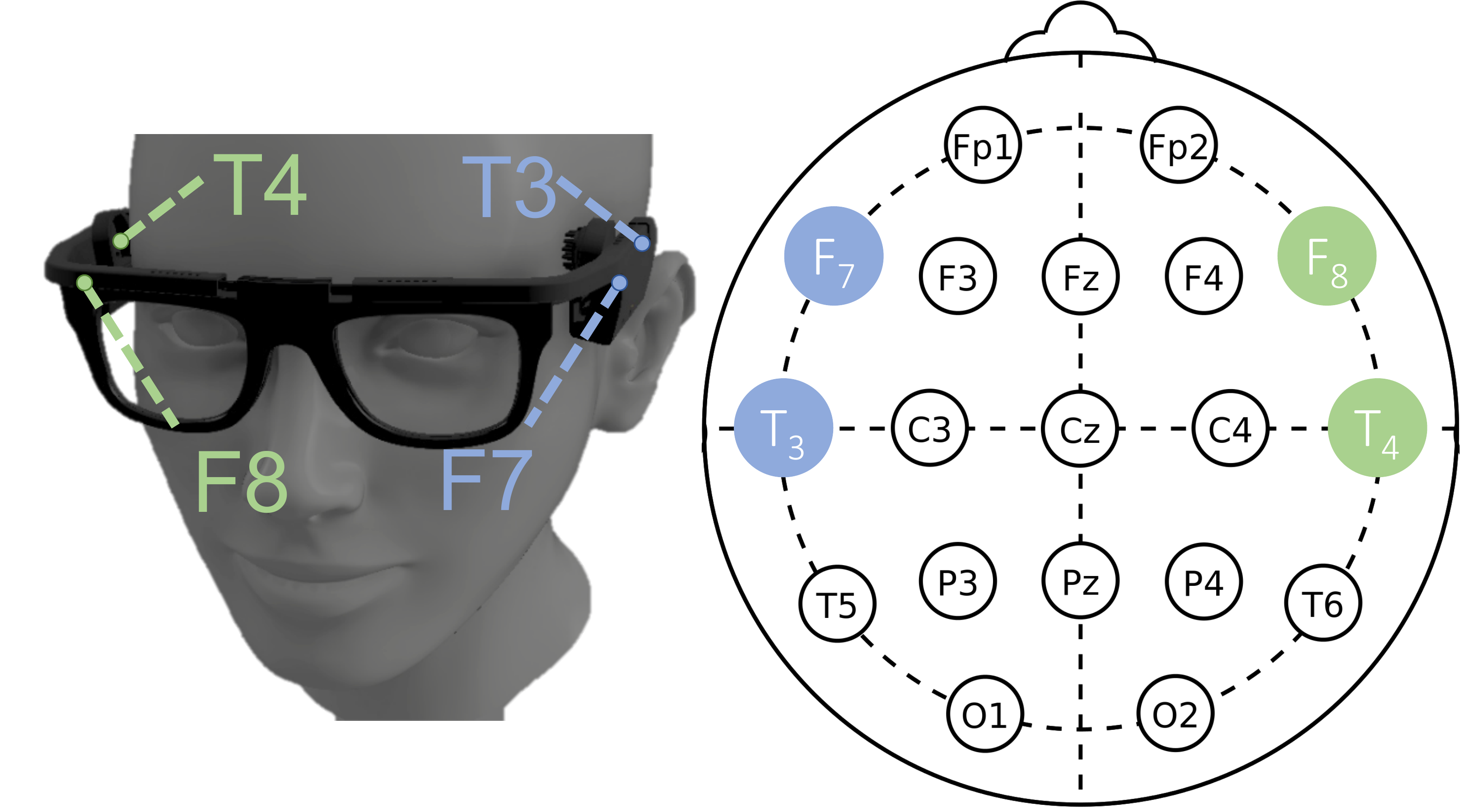}
		\caption{Electrodes locations of \textit{$F_7T_3$} and \textit{$F_8T_4$} for EEG monitoring using the e-Glass IoT wearable system for epileptic seizure detection \cite{sopic2018glass}.}  
		\label{eeg_electrodes}
	\end{center}
\end{figure}

In this previous context, for one of the 30 patients in the EPILEPSIAE dataset, the ECG and EEG signals had different lengths. As a result, it was impossible to synchronize them to label the data reliably. Thus, we have excluded the signals of this patient from our analysis and used the ECG and EEG data of 29 patients with 4603~hours of recordings containing 277 seizures. The data were acquired at a sampling rate of 256~Hz with 16-bit resolution. We employed both EEG and ECG to obtain the teacher network, but the student network only acquires the ECG signal. The EEG signals of the teacher network are from four electrodes (two channels): \textit{$F_7T_3$} and \textit{$F_8T_4$}, as shown in Fig~{\ref{eeg_electrodes}}. These electrodes are chosen so that the EEG signals can be acquired using e-Glass, an IoT wearable system based on these four EEG electrodes in a real-time scenario~{\cite{sopic2018glass}}. e-Glass is designed to be an inconspicuous system that helps patients to avoid the social stigma of wearing EEG head caps. In Section~{\ref{energy_consumption_analysis}}, we observe that it is also feasible to perform epileptic seizure detection with the teacher network in real-time using other edge AI systems (such as the Raspberry PI Zero~{\cite{gay2014raspberry}} or the recent PULP multi-core system~{\cite{de2020modular}}). However, due to the higher complexity of the teacher network compared to the student network, the teacher network consumes more energy, and it is unachievable to be deployed on such low-power edge AI platforms with limited resources. Conversely, the student network is less complex, making it more suitable for IoT wearable devices running for a longer period. Most importantly, the student network achieves the detection performance of the teacher network utilizing only the ECG signal and makes the signal acquisition easier for the patients.

We segmented each patient's seizure and non-seizure events into overlapping windows of 3 seconds and fed them into the proposed architecture. We split the dataset into training, validation, and test set. The training set contained 6,332 and 5,760,551 segments of seizures and non-seizures, respectively. Effective classification with imbalanced data is a crucial area of research, as high-class imbalance is naturally inherent in a real-world application such as epileptic seizure detection~\cite{beniczky2021machine}. To solve this issue, we perform undersampling. Undersampling means that from the majority class, which is non-seizure, we select as many segments as the minority class, which is the seizure. This selection maintains the probability distribution of the class during the training process.

Since we want to assess the separability between classes in a homogeneous way for different models and avoid the effect that different patients have different seizure frequencies, we considered a balanced scenario for both the validation and test set. Thus, we can evaluate the detection accuracy of the proposed framework in the classification of each segment separately. The validation set includes 400 segments of seizures and 400 segments of non-seizure. The test set consists of 784 segments of seizures and 784 segments of non-seizure. There are no segments from the same seizure or non-seizure in the training and test sets, simultaneously.

\subsection{Detection Performance Metrics}
To evaluate the detection performance of our proposed framework, we considered five different metrics. Sensitivity~(Sen) (Eq. \eqref{sen}) represents the percentage of ictal samples that are labeled correctly. Specificity (Spe) (Eq. \eqref{spe}) shows the percentage of inter-ictal samples that are labeled correctly. These metrics are defined as follows:

\begin{equation}
Sen = \frac{TP}{TP + FN},
\label{sen}
\end{equation}
\begin{equation}
Spe = \frac{TN}{FP + TN},
\label{spe}
\end{equation}
where TP, TN, FP and FN are true positive, true negative, false positive and false negative, respectively.

Geometric mean (Gmean) (Eq. \eqref{gmean})~\cite{fleming1986not} reflects both sensitivity and specificity, and measures the balance between classification performance in both classes. A low geometric mean indicates poor performance in classifying the seizure cases, even if the non-seizures cases are correctly classified.

\begin{equation}
Gmean = \sqrt{Sensitivity \times Specificity}.
\label{gmean}
\end{equation}

We also evaluate the accuracy (Acc) (Eq. \eqref{acc}), which represents the proportion of true positive results in the selected population.

\begin{equation}
Acc = \frac{TP + TN}{TP + TN + FP + FN}.
\label{acc}
\end{equation}

Finally, we use the F$_{1}$ score (Eq. \eqref{fone}), which is the harmonic mean of precision and recall. It gives a better measure of the incorrectly classified cases than the accuracy.

\begin{equation}
F_{1}\,score = \frac{2\;TP}{2\;TP + FP + FN}.
\label{fone}
\end{equation}

We also use a confusion matrix (Table~\ref{confusion_matrix_metric})~\cite{stehman1997selecting} to evaluate the detection performance of our proposed framework. Since the output can be one of two types of classes, the confusion matrix is one of the most intuitive and convenient to be used. In this case, the diagonal elements represent the number of cases for which the predicted label is equal to the correct label, while off-diagonal elements are those that are mislabeled by the classifier. The rows of the table are the ground truth classes, and the columns of the table are the classes of the predicted segment.

\begin{table}[h]
\caption{Confusion Matrix}
\label{confusion_matrix_metric}
\begin{center}
\begin{tabular}{c|cc}
\multicolumn{1}{c|}{} & 
\multicolumn{1}{c}{Seizure} & 
\multicolumn{1}{c}{Non-Seizure} \\ \hline
Seizure  & TP & FN   \\[1.5ex]
Non-Seizure  & FP   & TN \\ \hline
\end{tabular}
\end{center}
\end{table}

\subsection{Edge AI Evaluation Platform}
Wearable devices have small batteries and low-power processors compared to desktop processors. In this work, we use the Kendryte K210~\cite{kendryte_k210} and Raspberry Pi Zero~\cite{gay2014raspberry} to analyze and compare the energy consumption and timing requirements for continuous execution of the proposed approach. Note that the proposed approach must be executed repeatedly in real-time and have a satisfying detection performance. The Raspberry Pi Zero includes an ARM11 CPU running at 1 GHz, has 512MB RAM, and performs the inference process of a given DNN with power supplied via a micro USB connector. The Kendryte K210 is a chip system with specific circuits/components for machine vision and ML. This chip system employs advanced ultra-low processing with the help of a 64-bit dual-core processor equipped with a high-performance hardware
accelerator of the CNN. It supports convolution kernels, any form of activation function, and neural network parameter size up to 6 MB for real-time application.

\begin{figure}[t]
	\begin{center}
	\centering
		\includegraphics[scale=0.2]{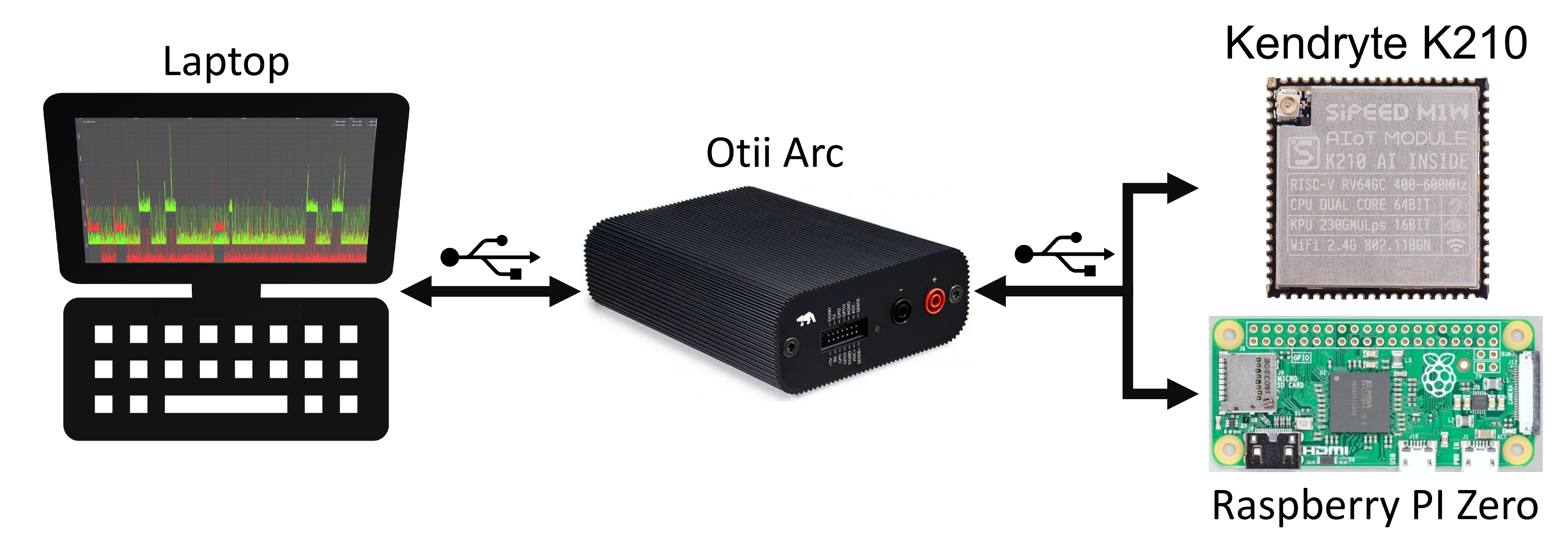}
		\caption{Hardware setup for the energy consumption measurement. Otii Arc is connected to the computer using a USB cable. The main output of the Otii Arc is connected to the voltage supply input of the IoT wearable platform. The current provided by the computer's USB port is used to power both the Otii Arc and edge AI platform. The Otii desktop application enables us to measure and analyze the energy consumption of the edge AI platform. In our evaluation, we considered a development system, which uses AI technologies embedded on a Kendryte K210 chip used in different IoT wearable systems \cite{kendryte_k210} or Raspberry Pi Zero. }
		\label{hw_setup}
	\end{center}
\end{figure}

We used Otii Arc~\cite{otii_arc} as a power analyzer and power supply for the inference process of our proposed approach. Otii Arc is a measurement tool for designing highly energy-efficient algorithms. It is powered via USB from the laptop and records both current and voltage, and it displays them in real-time for analysis and comparison. It provides up to 5 V output voltage and runs high-resolution current measurements with a sample rate up to 4 kHz for the range of 1 $\mu$A-5 A. Figure~\ref{hw_setup} shows the hardware setup of our measurement.

We considered the Kendryte K210 chip and Raspberry Pi Zero as they have comparable processing capabilities to modern wearable architectures~{\cite{pullini2019mr}}. We also discuss the potential benefit of using PULP-based ultra-low-power platforms and architectures proposed in~{\cite{de2020modular}} for wearable biomedical systems to further reduce the power consumption.

\subsection{Learning Parameters}
We trained our proposed networks from scratch using pre-processed 3-second ECG and EEG segments. The weights initialization of the layers follows a normal distribution with zero mean and 0.01 as the standard deviation. We initialize all the biases to zero. During the training, the network learns the correlation between the input and the output consisting of two nodes and adjusts the parameters of the model to minimize the cross-entropy loss. For the binary classification, the final output of DNN can have a single output and a threshold, or we can use a multi-class classification with only two nodes, so each class gets its output neuron. However, the two-node outputs technique code is the same for the multi-class classification problem and can be easily extended to multi-class classification in future work. Finally, we use the Adam optimizer~\cite{kingma2014adam} with a mini-batch size of 16 across all patients. The base learning rate is 10$^{-5}$, and the DNNs are trained for up to $10^4$ iterations and are implemented on Tensorflow 1.14.0~\cite{tensorflow2015-whitepaper}.

\section{Evaluation}
\label{sec:Result}

In this section, we present the assessment of the seizure detection performance and energy consumption of our proposed knowledge distillation approach on the Kendryte K210 and the Raspberry Pi Zero unit.

\subsection{Detection Performance Analysis}
\label{detection_performance_analysis}

\begin{figure*}[ht]%
    \begin{center}
	\centering
		\includegraphics[scale=0.45]{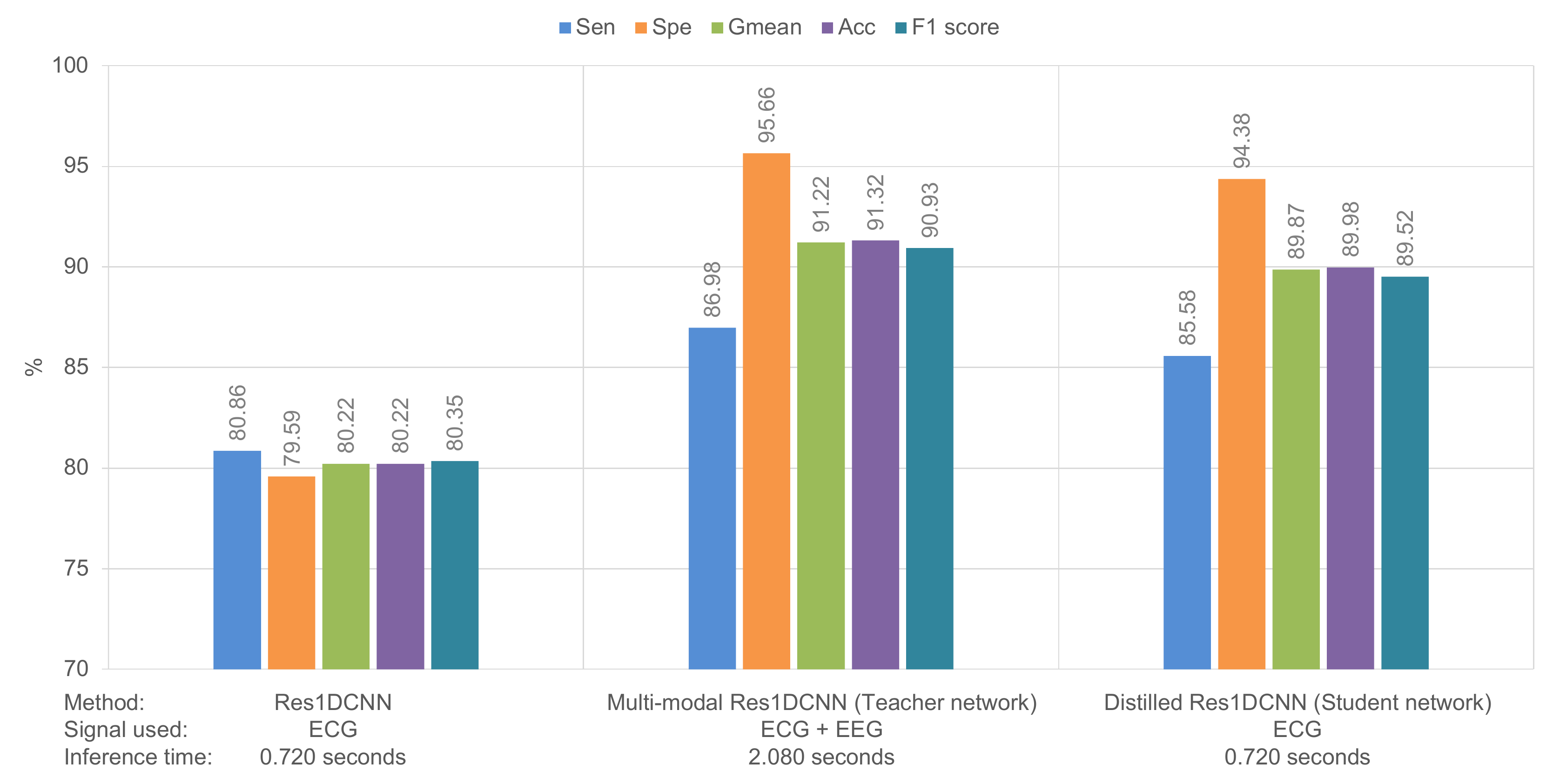}
		\caption{Comparison of Res1DCNN in epileptic seizure detection versus multi-modal Res1DCNN and distilled Res1DCNN. It shows that distilling performs well for transferring the knowledge from an ensemble DNN (Multi-modal Res1DCNN using ECG + EEG) into a smaller, distilled DNN (Distilled Res1DCNN using ECG).}
		\label{chart_comparison}
	\end{center}
\end{figure*}

Figure~\ref{chart_comparison} shows the effects of epileptic seizure detection of the proposed multi-modal Res1DCNN (Teacher network) using both ECG and EEG segments, and it shows the comparison with the Res1DCNN where we employ only ECG segments. We observe that our proposed multi-modal Res1DCNN achieves the geometric mean and F$_{1}$ score of 91.22\% and 90.93\%, which are in line with other state-of-the-art results based on CNNs applied to EEG signals. For example, in~\cite{gabeff2021interpreting} the authors used four EEG channels and achieved F$_{1}$ score of 87.30\%. The multi-modal Res1DCNN outperforms the Res1DCNN trained exclusively on ECG by 11.00\% in terms of the \emph{Gmean}. It shows that extracting additional features from EEG signals is more effective in seizure detection. However, the size of the DNN model is limited by the memory capacity of the embedded medical platform. The memory capacity bottleneck limits the practical use of large DNNs~\cite{diamos2016persistent}, such as multi-modal Res1DCNN. Then, knowledge distillation acts as a compression method to reduce the hardware footprint of a DNN model to decrease its inference latency without overtly affecting inference accuracy. Figure~\ref{chart_comparison} evaluates the detection performance and inference time of our proposed knowledge distillation framework applied to multi-modal Res1DCNN. We show that the proposed scheme achieves the desired effect of distilling an ensemble of DNN models such as multi-modal Res1DCNN into a single DNN model that works significantly better than a DNN model with the same size learned directly using the same trained data. In particular, we observe that multi-modal Res1DCNN achieves a geometric mean of 91.22\% when the training and inference are performed using both ECG and EEG signals. In contrast, in distilled Res1DCNN (student network), which is trained directly on the ECG, this value decreases by only 1.35\%.

To understand the trade-off between the training efficiency and model detection accuracy, in Table~\ref{confusion_matrix_comparison}, we compare the confusion matrices of the DNNs when they are trained using only ECG segments without knowledge distillation versus with knowledge distillation and multi-modal DNN, which requires both ECG and EEG segments. The detection accuracy using the distilled DNN is slightly lower than when the training process is performed using a more complex multi-modal DNN. In particular, in multi-modal Res1DCNN, we detect 86.99\% of epileptic seizure segments, while in the distilled Res1DCNN, this value decreases to 85.59\%. The distillation approach mainly helps reduce the number of false positives from 160 to 44.


\begin{table}[h]
\caption{Comparison of the confusion matrices of Res1DCNN, multi-modal Res1DCNN, and distilled Res1DCNN in epileptic seizure detection.}
\label{confusion_matrix_comparison}
\begin{tabular}{c|cc}
\multicolumn{3}{c}{$\qquad\qquad\qquad$ Number of segments}\\
\multicolumn{1}{c|}{} & 
\multicolumn{1}{c}{Seizure} & 
\multicolumn{1}{c}{Non-Seizure} \\ \hline
Seizure  & 634 & 150   \\[1.5ex]
Non-Seizure  & 160  & 624 \\ \hline
\end{tabular}
\begin{tabular}{c|cc}
\multicolumn{3}{c}{$\qquad$Normalized}\\
\multicolumn{1}{c|}{} & 
\multicolumn{1}{c}{Seizure} & 
\multicolumn{1}{c}{Non-Seizure} \\ \hline
  & 80.87\% & 19.13\%   \\[1.5ex]
  & 20.41\%   & 79.59\% \\ \hline
\end{tabular}
\begin{center}
    \textbf{Res1DCNN}
\end{center}
\vspace{0.2cm}
\begin{tabular}{c|cc}
\multicolumn{3}{c}{$\qquad\qquad\qquad$ Number of segments}\\
\multicolumn{1}{c|}{} & 
\multicolumn{1}{c}{Seizure} & 
\multicolumn{1}{c}{Non-Seizure} \\ \hline
Seizure  & 682 & 102   \\[1.5ex]
Non-Seizure  & 34   & 750 \\ \hline
\end{tabular}
\begin{tabular}{c|cc}
\multicolumn{3}{c}{$\qquad$Normalized}\\
\multicolumn{1}{c|}{} & 
\multicolumn{1}{c}{Seizure} & 
\multicolumn{1}{c}{Non-Seizure} \\ \hline
  & 86.99\% & 13.01\%   \\[1.5ex]
  & 4.34\%   & 95.66\% \\ \hline
\end{tabular}
\begin{center}
    \textbf{Multi-modal Res1DCNN (Teacher network)}
\end{center}
\vspace{0.2cm}
\begin{tabular}{c|cc}
\multicolumn{3}{c}{$\qquad\qquad\qquad$ Number of segments}\\
\multicolumn{1}{c|}{} & 
\multicolumn{1}{c}{Seizure} & 
\multicolumn{1}{c}{Non-Seizure} \\ \hline
Seizure  & 671 & 113   \\[1.5ex]
Non-Seizure  & 44   & 740 \\ \hline
\end{tabular}
\begin{tabular}{c|cc}
\multicolumn{3}{c}{$\qquad$Normalized}\\
\multicolumn{1}{c|}{} & 
\multicolumn{1}{c}{Seizure} & 
\multicolumn{1}{c}{Non-Seizure} \\ \hline
  & 85.59\% & 14.41\%   \\[1.5ex]
  & 5.61\%   & 94.39\% \\ \hline
\end{tabular}
\begin{center}
    \textbf{Distilled Res1DCNN (Student network)}
\end{center}
\end{table}

\subsection{Alternatives Scenarios for Detection Performance Analysis}
\label{detection_performance_analysis_alternatives_scenarios}
We also considered two different scenarios. The first one is when the teacher network does not use three identical Res1DCNN for each EEG1, EEG2, and ECG signal. Instead, it uses only one DNN that receives an input of three channels. Therefore, we modified the first layer of Res1DCNN, so it accepts an input of three channels (EEG1, EEG2, and ECG). We observed that the teacher network's detection accuracy (sensitivity and specificity) with only one modified Res1DCNN is lower (by 3.18\% and 1.66\%) than when the training process is performed using a more complex teacher model consisting of three Res1DCNN. Conversely, the teacher model with one modified Res1DCNN consumes less energy. However, in this work, since our goal is to develop high-precision and low-power wearable systems (student model) using a single-biosignal input (ECG signal), we have designed a more complex teacher model. It relies on independent 1-dimensional networks for each biosignal to maximize the seizure detection performance.

Another scenario we have considered is when the student network uses only EEG signals (EEG1 and EEG2). We observed the detection accuracy (sensitivity and specificity) using a more complex distilled DNN (student model) including EEG1 and EEG2 signals is slightly higher (by 0.51\% and 0.9\%) than when the training process is performed using only ECG. However, EEG testing comes with certain adverse conditions. Wearing EEG head caps causes social stigma and discomfort for the patients. To overcome this issue, in this work, we proposed a framework to perform the epileptic seizure detection using only the ECG signal while achieving the detection accuracy of a system where both the ECG and EEG signals are utilized. The ECG signal acquisition demands less energy, and acquisition devices are widely accessible worldwide. As a result, more patients would benefit from long-term monitoring for epileptic seizure detection.

\subsection{Energy Consumption Analysis}
\label{energy_consumption_analysis}

A key challenge for low-energy embedded medical platforms with limited computational resources is designing and implementing an epileptic seizure detection algorithm based on DNNs for long-term patient monitoring. For instance, the e-Glass wearable system~\cite{sopic2018glass} shown in Fig.~\ref{eeg_electrodes} contains a 570 mAh battery and features an ultra-low-power 32-bit microcontroller STM32L151~\cite{STM32L151} with an ARM\textsuperscript{\textregistered} Cortex\textsuperscript{\textregistered}-M3 with 48 KB RAM and 384 KB Flash. In the case of an epileptic seizure, e-Glass communicates with a smartphone or smartwatch using Bluetooth low energy (nRF8001)~\cite{nRF8001} and sends a warning to the caregivers. 


In this paper, to analyze the complexity, lifetime, and energy efficiency of our approach, we consider the Kendryte K210 and Raspberry Pi Zero platforms. 
In the implementation code, all the computations and storage are in 16-bit fixed-point. We chose 13 bits for the fractional part using the results of the validation set. We observed that dedicating more bits to the fractional part causes overflows in the computations. On the other hand, reducing the number of fraction bits gives rise to a considerable accuracy drop. Since we use the fixed-point representation of numbers, we save the amount of storage by a factor of 4 compared to 64-bit floating-point operations. This compression is crucial because it enables our network to be applicable on various memory-limited embedded devices. However, the total accuracy of the model is reduced by 0.9\% because of the quantization in 16-bit.

Table~\ref{rune_time} shows the seizure detection execution time per 3-second segment for each DNN. The represented numbers are obtained by running the experiments for the whole test set, including 1568 segments of 3-second segments. We observe that due to fewer parameters, the proposed knowledge distillation network, which requires only the ECG signal, runs 2.9 times faster than the multi-modal Res1DCNN on the Raspberry Pi Zero. At the same time, the proposed network achieves a detection performance comparable to multi-modal Res1DCNN (see Fig.~\ref{chart_comparison}). In addition to the reduced memory and computational burden, the most relevant advantage of the proposed distilled version is that it avoids the acquisition and process of EEG data. We also observe that the network's end-to-end response time for a 3-second segment is only 720 milliseconds on the Raspberry Pi Zero and 1,040 milliseconds on the Kendryte K210; thus, we can continuously monitor the patients in real-time.

\begin{table}[h]
\caption{Quantitative evaluation results of run-time for every 3-second segment in different network architectures.}

\label{rune_time}
\begin{tabular}{|c|c|c|}
\hline
\textbf{Method}                                                                                   & \textbf{Platform}                  & \textbf{Run time (millisec.)} \\ \hline \hline
Res1DCNN                                                                                          & Raspberry Pi Zero                  & 720.15 $\pm$ 32.46                             \\ \hline
\multirow{2}{*}{\begin{tabular}[c]{@{}c@{}}Multi-modal Res1DCNN\\ (Teacher network)\end{tabular}} & \multirow{2}{*}{Raspberry Pi Zero} & \multirow{2}{*}{2,080.56 $\pm$ 12.56}            \\
                                                                                                  &                                    &                               \\ \hline
\begin{tabular}[c]{@{}c@{}}Distilled Res1DCNN\\ (Student network)\end{tabular}                    & Raspberry Pi Zero                  & 720.15 $\pm$ 32.46                             \\ \hline
\begin{tabular}[c]{@{}c@{}}Distilled Res1DCNN\\ (Student network)\end{tabular}                    & Kendryte K210                      & 1,040.64 $\pm$ 5.67                             \\ \hline
\end{tabular}
\end{table}

\begin{figure}[t]
     \centering
     \begin{subfigure}
         \centering
         \includegraphics[scale=0.33]{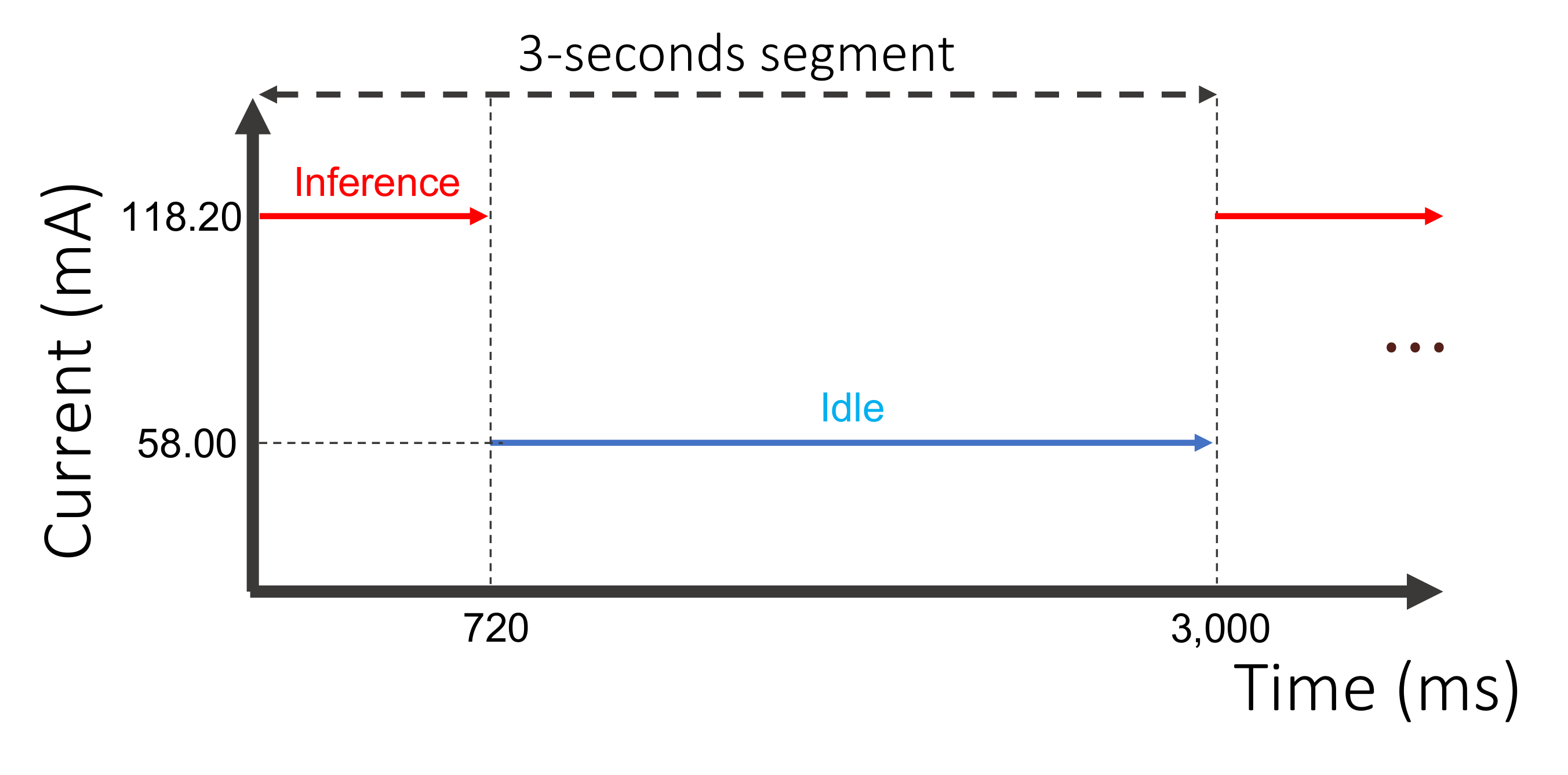}
         \caption{Real-time energy consumption monitoring on Raspberry Pi Zero. Raspberry Pi Zero performs epileptic seizure detection using Distilled Res1DCNN of a 3-seconds segment in 720 milliseconds and then goes to idle mode.}
         \label{real_time_energy_cons}
     \end{subfigure}
     \hfill
     \begin{subfigure}
         \centering
         \includegraphics[scale=0.33]{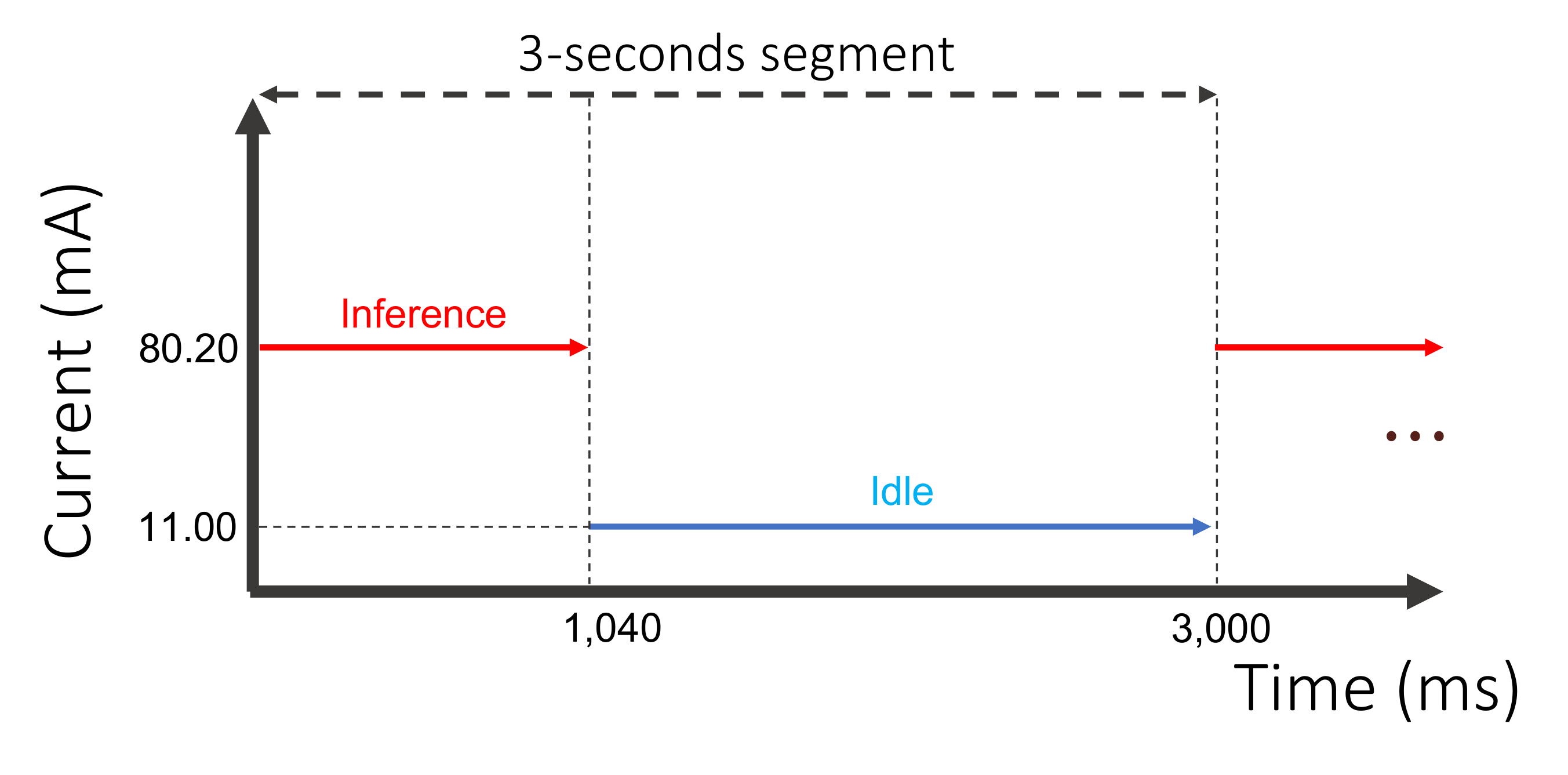}
         \caption{Real-time energy consumption monitoring on the Kendryte K210. The Kendryte K210 performs epileptic seizure detection using Distilled Res1DCNN of a 3-seconds segment in 1,036 milliseconds and then goes to idle mode with considerably lower energy consumption.}
         \label{real_time_energy_cons_grove_ai}
     \end{subfigure}
     \hfill
\end{figure}

Epilepsy is characterized by unpredictable seizures and can cause other health problems; thus, the patients have to be monitored on a long-term basis. Table~\ref{long_term_monitor} evaluates the battery lifetime of our proposed distilled network using the battery of the e-Glass~\cite{sopic2018glass}, which is 570 mAh. We observe that our proposed knowledge distillation model operates for 7.86 hours on the Raspberry Pi Zero and 16.29 hours on the Kendryte K210 on a single charge to perform real-time epileptic seizure detection. The proposed distilled network achieves a 37.65\% energy reduction sacrificing just 1.5\% of the accuracy. The battery life is measured considering the real-time energy consumption shown in Fig.~\ref{real_time_energy_cons} and ~\ref{real_time_energy_cons_grove_ai}. We observe that the Raspberry Pi Zero executes the inference of the 3-seconds segment in 720~ms, then goes to idle mode and waits for the next 3-seconds segment. As Fig.~\ref{real_time_energy_cons} shows, the Raspberry Pi Zero consumes a considerable amount of energy in idle mode. Therefore, as shown in Fig. ~\ref{real_time_energy_cons_grove_ai} we considered other wearable platforms such as the Kendryte K210 that consume less power in idle mode which would be very beneficial for the patients to perform long-term monitoring. We can improve the battery life of the edge device by using the PULP-based ultra-low-power wearable platform proposed in~{\cite{de2020modular}}, which consumes only 0.76 mA when the system is clock gated with respect to 23.58 mA when the system is running at 110 MHz, at the lowest energy point of the platform 0.8 V. In this scenario, assuming that the inference time will be similar for the PULP after parallelization~{\cite{9560136}}, we can achieve a battery life of 91.33 hours with the same battery capacity.

\begin{table}[h]
\caption{Battery life of an edge device using the e-Glass~\cite{sopic2018glass} battery for running each network architecture to perform patient monitoring.}

\label{long_term_monitor}
\begin{tabular}{|c|c|c|}
\hline
\textbf{Method}                                                                                   & \textbf{Platform}                  & \textbf{Battery life (hours)} \\ \hline \hline
Res1DCNN                                                                                          & Raspberry Pi Zero                  & 7.86 $\pm$ 0.09                             \\ \hline
\multirow{2}{*}{\begin{tabular}[c]{@{}c@{}}Multi-modal Res1DCNN\\ (Teacher network)\end{tabular}} & \multirow{2}{*}{Raspberry Pi Zero} & \multirow{2}{*}{5.71 $\pm$ 0.01}            \\
                                                                                                  &                                    &                               \\ \hline
\begin{tabular}[c]{@{}c@{}}Distilled Res1DCNN\\ (Student network)\end{tabular}                    & Raspberry Pi Zero                  & 7.86 $\pm$ 0.09                             \\ \hline
\begin{tabular}[c]{@{}c@{}}Distilled Res1DCNN\\ (Student network)\end{tabular}                    & Kendryte K210                      & 16.29 $\pm$ 0.06                           \\ \hline
\end{tabular}
\end{table}

\section{Conclusion}
\label{sec:Conclusion}

The development of IoT wearable systems that can accurately detect complex pathologies, such as brain disorders, in long term and with minimal discomfort is still an open challenge. In this work, we have proposed a new knowledge distillation approach to develop high-precision and low-power IoT wearable systems using single-biosignal input for epileptic seizure detection. 
As the starting point, for our teacher network, we have designed a multi-modal DNN, using information from both ECG and EEG signals, that relies on independent 1-dimensional networks for each biosignal to maximize the detection performance. Then, we have used knowledge distillation to develop a compressed student network that relies exclusively on ECG data during run-time operation of the IoT wearable system. Besides reducing the energy consumption, moving from multi-biosignal to single-biosignal has resulted in removing other major drawbacks today in IoT wearable devices when they need to be deployed in real-life setups, such as discomfort, stigma, and synchronization problems where multiple biosignals need to be acquired. Indeed, these benefits are achieved for our distilled network design considering IoT setups, while achieving a comparable detection performance with respect to the multi-modal teacher DNN system. The results of our approach implemented on different edge AI platforms for the IoT wearable context, and tested on the EPILEPSIAE dataset, have shown a 37.65\% reduction in energy consumption. In comparison, the resulted sensitivity and specificity are only 1.5\% and 1.3\% lower than in the initial multi-modal DNN teacher system. Thus, our proposed approach is ideal for the development of IoT wearable setups as it removes the burden of acquiring and synchronizing multiple devices to make valid medical assessments.
\printbibliography
\end{document}